\documentclass[%
 aip,
rsi,%
amsmath,amssymb,
 reprint,%
longbibliography,
]{revtex4-1}

\usepackage{graphicx}
\usepackage{dcolumn}
\usepackage{bm}

\begin{document}

\preprint{AIP/123-QED}

\title[Three-dimensional imaging-genetics of the heart]{Three-dimensional Cardiovascular Imaging-Genetics:\\ A Mass Univariate Framework}

\author{Carlo Biffi} 
\affiliation{Department of Computing, Imperial College London, South Kensington Campus, London, UK.}
\affiliation{MRC London Institute of Medical Sciences, Faculty of Medicine, Imperial College London, Hammersmith Hospital Campus, London, UK.}
\author{Antonio de Marvao} 
\affiliation{MRC London Institute of Medical Sciences, Faculty of Medicine, Imperial College London, Hammersmith Hospital Campus, London, UK.}
\author{Mark I. Attard} 
\affiliation{MRC London Institute of Medical Sciences, Faculty of Medicine, Imperial College London, Hammersmith Hospital Campus, London, UK.}
\author{Timothy J.W. Dawes} 
\affiliation{MRC London Institute of Medical Sciences, Faculty of Medicine, Imperial College London, Hammersmith Hospital Campus, London, UK.}
\author{Nicola Whiffin} 
\affiliation{MRC London Institute of Medical Sciences, Faculty of Medicine, Imperial College London, Hammersmith Hospital Campus, London, UK.}
\affiliation{National Heart and Lung Institute, Imperial College London, London, UK.}
\affiliation{NIHR Royal Brompton Cardiovascular Biomedical Research Unit, London, UK.}
\author{Wenjia Bai} 
\affiliation{Department of Computing, Imperial College London, South Kensington Campus, London, UK.}
\author{Wenzhe Shi} 
\affiliation{Department of Computing, Imperial College London, South Kensington Campus, London, UK.}
\author{Catherine Francis}
\affiliation{National Heart and Lung Institute, Imperial College London, London, UK.}
\affiliation{NIHR Royal Brompton Cardiovascular Biomedical Research Unit, London, UK.}
\author{Hannah Meyer} 
\affiliation{European Molecular Biology Laboratory, European Bioinformatics Institute, Hinxton, UK.}
\author{Rachel Buchan} 
\affiliation{National Heart and Lung Institute, Imperial College London, London, UK.}
\affiliation{NIHR Royal Brompton Cardiovascular Biomedical Research Unit, London, UK.}
\author{Stuart A. Cook}
\affiliation{MRC London Institute of Medical Sciences, Faculty of Medicine, Imperial College London, Hammersmith Hospital Campus, London, UK.}
\affiliation{National Heart and Lung Institute, Imperial College London, London, UK.}
\affiliation{NIHR Royal Brompton Cardiovascular Biomedical Research Unit, London, UK.}
\affiliation{National Heart Centre Singapore, Singapore.}
\affiliation{Duke National University Singapore, Singapore.}
\author{Daniel Rueckert} 
\affiliation{Department of Computing, Imperial College London, South Kensington Campus, London, UK.}
\author{Declan P. O'Regan} 
\affiliation{MRC London Institute of Medical Sciences, Faculty of Medicine, Imperial College London, Hammersmith Hospital Campus, London, UK.}

\begin{abstract}
\textbf{Background}: Left ventricular (LV) hypertrophy is a strong predictor of cardiovascular outcomes, but its genetic regulation remains largely unexplained. Conventional phenotyping relies on manual calculation of LV mass and wall thickness, but advanced cardiac image analysis presents an opportunity for high-throughput mapping of genotype-phenotype associations in three dimensions (3D).
\textbf{Methods}: High-resolution cardiac magnetic resonance images were automatically segmented in 1,124 healthy volunteers to create a 3D shape model of the heart. Mass univariate regression was used to plot a 3D effect-size map for the association between wall thickness and a set of predictors at each vertex in the mesh. The vertices where a significant effect exists were determined by applying threshold-free cluster enhancement to boost areas of signal with spatial contiguity. \textbf{Results}: Experiments on simulated phenotypic signals and SNP replication show that this approach offers a substantial gain in statistical power for cardiac genotype-phenotype associations while providing good control of the false discovery rate. \textbf{Conclusion}: This framework models the effects of genetic variation throughout the heart and can be automatically applied to large population cohorts. \textbf{Availability}: The proposed approach has been coded in an R package freely available at \href{https://doi.org/10.5281/zenodo.834610}{https://doi.org/10.5281/zenodo.834610} together with the clinical data used in this work.
\end{abstract}


\maketitle
\noindent

\section{Introduction}
\begin{figure}
\includegraphics[width=0.45\textwidth]{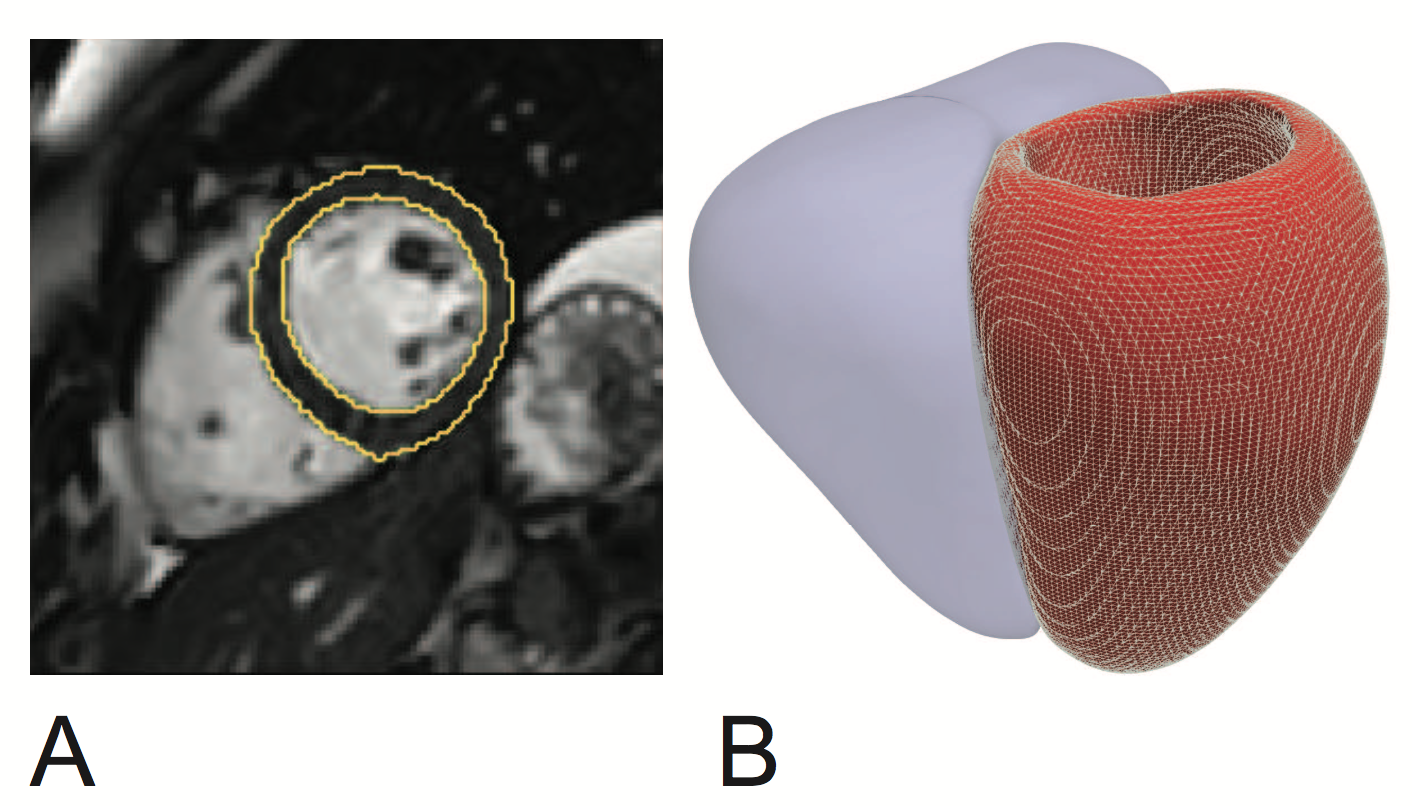}
\caption{\label{fig1} {\bf Computational image analysis.} (A) Short axis cardiac magnetic resonance image demonstrating automated segmentation of the endocardial and epicardial boundaries of the left ventricle. (B) The segmentation is used to construct a three dimensional mesh of the cardiac surfaces (left ventricle shown in red, right ventricle  in blue) that is co-registered to a standard coordinate space. Phenotypic parameters, such as wall thickness, are then derived for each vertex in the model.}
\end{figure}

One of the most complex unanswered questions in cardiovascular biology is how genetic and environmental factors influence the structure and function of the heart as a three-dimensional (3D) structure \citep{li2016transcriptomic}. This is relevant for understanding the penetrance and expressivity of variants associated with inherited cardiac conditions as well as exploring the biology of heart development and within-population variation. Cardiac magnetic resonance (CMR) is the gold-standard for quantitative imaging \citep{hundley2010accf}, providing a rich source of anatomic and motion-based data, however conventional phenotyping relies on manual analysis reducing the variables of interest to global volumes and mass. Computational image analysis, by which machine learning is used to annotate and segment the images, is gaining traction as a means of representing detailed 3D phenotypic variation at thousands of vertices in a standardized coordinate space (FIG. \ref{fig1}) \citep{young2009,young_atlas}. \\
One approach to inference is to transform the spatially-correlated data into a smaller number of uncorrelated principal components \citep{medrano2015}, however these modes would not provide an explicit model relating genotype to phenotype. A more powerful approach may be to derive a statistic expressing evidence of a given effect at each vertex of the 3D model, hence creating a so-called statistical parametric map - a concept widely used in functional neuroimaging \citep{penny2011}. In this paper we extend techniques developed in the neuroscience domain to cardiovascular imaging-genetics by implementing a mass univariate framework to map associations between genetic variation and a 3D phenotype. Such an approach would provide overly-conservative inferences without considering spatial dependencies in the underlying data and so we validated the translation of threshold-free cluster-enhancement (TFCE) to cardiovascular phenotypes for the sensitive detection of coherent signal \citep{smith2009} as well as implementing robust control for multiple testing. The feasibility of the proposed methodology to derive computationally-efficient inferences on imaging-genetics datasets has been tested through experiments on clinical and synthetic data using an R package developed for this purpose. 

\section{Methods}

\subsection{Study population}
The healthy volunteers dataset used in this study is part of the UK Digital Heart Project at Imperial College London \citep{bai2015} (see \emph{Appendix A} for full cohort characteristic and acquisition details). To capture the whole-heart phenotype, a high-spatial resolution 3D balanced steady-state free precession cine sequence was performed on a 1.5-T Philips Achieva system (Best, the Netherlands). Images were stored on an open-source database (MRIdb, Imperial College London, UK) \citep{mridb}. Conventional volumetric analysis of the cine images was performed using CMRtools (Cardiovascular Imaging Solutions, London, UK) following a standard protocol \citep{schulz2013standardized}. \\
Genotyping of common variants was carried out using an Illumina HumanOmniExpress-12v1-1 single nucleotide polymorphism (SNP) array (Sanger Institute, Cambridge). Clustering, calling and scoring of SNPs was performed using Illumina GenCall software. Samples were pre-phased with SHAPEIT \citep{delaneau2013improved} and imputation was performed using IMPUTE2 \citep{howie2009flexible} with the UK10K dataset as a reference (www.uk10k.org). Quality of the genotypes was evaluated both on a per-individual and per-marker level using in-house Perl scripts. SNPs were removed if they had a Impute Information (INFO) score $<0.4$, missing call rate in  more  than  1$\%$ of samples, minor allele frequency of less than 1$\%$ or deviated significantly from Hardy-Weinberg  equilibrium ($p>0.001$). Only non-related individuals with "CEU" ethnicity were retained. The total genotyping rate in these individuals was 0.997 and the total number of variants available was 9.4 million.

\subsection{Atlas-based segmentation and co-registration}
All image processing was performed with Matlab (MathWorks, Natick, Mass). A validated cardiac segmentation and co-registration framework was used which has previously been described in detail \citep{bai2015,deMarvao2015}. A 3D shape model (at end-diastole and end-systole) was created encoding phenotypic variation in our study population at 49,876 epicardial vertices and visualised in a standard coordinate space (FIG. \ref{fig1}) \citep{bai2015}. Wall thickness (WT) was measured by computing the distance between respective vertices on the endocardial and epicardial surfaces at end-diastole.

\begin{figure*}
\includegraphics[width=\textwidth]{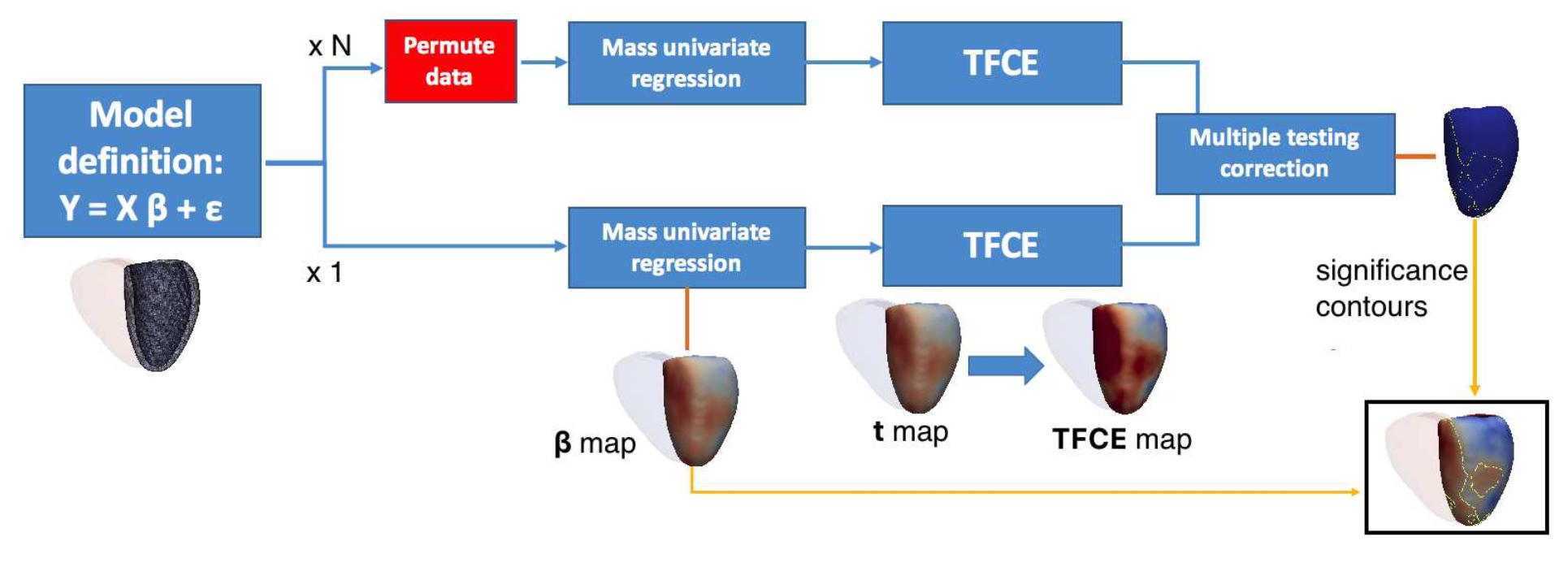}
\caption{{\bf Outline of three-dimensional mass univariate framework.} A statistical atlas provides point-wise measures of ventricular geometry and function which can be linked to a given predictor through a general linear model. Using mass univariate regression, three-dimensional maps of a test statistic and the degree of association ($\beta$) can be derived. Threshold free cluster enhancement (TFCE) coupled with permutation testing produces vertex-wise p-values weighted to the degree of coherent spatial support. Finally, p-values are corrected for multiple testing. Regression coefficients enclosed by significance contours are represented on a model of the left ventricle.}
\label{fig2}
\end{figure*}

\subsection{Overview of the approach}
In the following sections we introduce a framework for deriving associations between clinical/genetic parameters and a 3D cardiac phenotype which is outlined in FIG. \ref{fig2}. Briefly, a general linear model is fitted at each vertex to extract the regression coefficient associated with the variable of interest (mass univariate regression). Threshold-free cluster enhancement (TFCE) is then applied to boost belief in extended areas of coherent signal in the derived vertex-wise statistical map.  The points where a significant effect exists are determined by applying TFCE on the obtained t-statistic map and on $N$ t-statistic maps obtained through permutation testing, derived under the null hypothesis of no effect. Then, at each vertex the frequentist probability of having obtained a higher TFCE score by chance is regarded as the p-value related to the regression coefficient $\beta$. Finally, the derived p-values are adjusted for multiple testing. The permutation testing procedure employed by this approach is the Freedman-Lane procedure \citep{freedman1983}, whilst a false discovery rate (FDR) correction using the Benjamini-Hochberg procedure \citep{benjamini1995} is applied to correct for multiple testing. 

\subsection{Mass univariate analysis}
The association between a ventricular phenotype mapped onto a 3D mesh and one or more clinical variables can be described using a general linear model of the form $Y = \beta X + \epsilon$, where $Y$ is a vector of, for example, WT values at each vertex and $X$ is a design matrix that can be used to model the effect of interest and contains in each column the subject's values of clinical co-variates as well as the intercept term. These variables can be numerical (such as age or weight), categorical or expressing interaction between them. In particular, categorical variables can be exploited to express either different categories (such as gender or ethnicity) or the presence/absence in a binary form of a clinical condition (such as the presence of genetic mutation or a specific disease). $\beta$ is the regression coefficient vector to be estimated and $\epsilon$ represents the noise or error term, which is assumed to be a zero-mean Gaussian process and represents the variability of $Y$ not explained by the model. The regression coefficient can be standardized by normalizing to mean 0 and unit-variance the columns of $X$ and $Y$. As a result, $\beta$ will represent the amount of variation of $Y$ in units of its standard deviation when $X$ is increased by one standard deviation, allowing comparisons between variables. \\ 
The same model can be fitted at each ventricular vertex independently (mass univariate regression) and statistics can be extracted and corrected for multiple testing in order to test one or more statistical hypotheses. In a parametric framework, the t-statistic computed as $t=\frac{\beta}{s.e.(\beta)}$ is typically used in the neuroimaging literature to contrast the null hypothesis $H_0: \beta = 0$ (no association between the predictor and the phenotype under study), where $s.e.(\beta)$ is the standard error of the estimator of $\beta$ \citep{penny2011}. The regression coefficients $\beta$ and their related p-value thus obtained can be plotted to display, at high resolution on the whole 3D ventricle, the magnitude and spatial distribution of a given association. However, this approach underestimates associations where the signal is more spatially correlated than noise coherence. For this reason non-parametric statistics such as TFCE are valuable to increase the statistical power of the approach.

\subsection{Threshold-free cluster enhancement on a cardiac atlas}The value of a statistic h obtained through mass univariate regression at a vertex p - a t-statistic in our context - is transformed by TFCE using the following integral:

\begin{equation}
 TFCE(p) = \int_{h=0}^{h_p} e(h)^E \; h^H \; \delta h \simeq \sum_{h=0}^{h_p}  e(h)^E \; h^H \; \Delta h
\label{eq1}
\end{equation}

\noindent In the equation $h_p$ is the value of the vertex statistic, $e(h)$ is the extent of the cluster with cluster-forming threshold $h$ that contains $p$, and $E$ and $H$ are two parameters usually set to 0.5 and 2 for empirical and analytical motivations \citep{smith2009}. In computational algorithms the integral is estimated using a discrete sum. The computational model of the heart is defined as a 3D mesh composed of non-congruent triangles where at each vertex pointwise phenotypic variables are stored for each subject. The translation of TFCE to a cardiac model is not straightforward as the model is not composed of a regular grid of voxels (as in brain imaging applications) but instead forms a mesh of vertices. We addressed this problem by associating an area to each vertex \emph{i} equivalent to the mesh area closest to that vertex. In computing Eq. \ref{eq1} at each vertex, the most time consuming part is deriving $e(h)$ - the area of all the elements connected to p that have a statistic value greater or equal to $h$ - as a different $e(h)$ needs to be computed for each vertex of the mesh and for each term of the sum. However, the TFCE score associated to a vertex $p$ of a specific $h$ in the summation consists of the same score that should be associated with all the vertices which contribute to $e(h)$. Therefore, the computational time of the TFCE method can be significantly reduced by sampling the interval between the maximum and minimum statistic $h$ in the statistic map so as to use each sampled value as a threshold $\tilde{h}$ for the selection of vertices with a greater statistic value in the case of positive threshold, or less than $\tilde{h}$ in the case of a negative threshold. The edges of the graph are defined from a list containing the nearest neighbours of each vertex, and which is filtered at each iteration to contain only the vertices selected by the thresholding operation, resulting in one or more graphs of connected vertices. In this way, all the possible patterns of signal on the ventricle can be discovered without relying on assumptions about the geometry of cluster shapes. For all the obtained graphs including more than two vertices the TFCE score can be computed and associated to all the vertices that belong to them. The final TFCE score is the sum of all the TFCE scores thus obtained.

\subsection{Permutation testing}
The p-value associated with the regression coefficient computed at each atlas vertex can be derived via permutation testing. In particular, by permuting $N$ times the input data, $N$ TFCE scores maps can be obtained. It is important that the adopted permutation strategy guarantees the exchangeability assumption, \emph{i.e.} permutations of $Y$ given $ X$ do not alter the joint distribution of the dependent variables under the null hypothesis. In the proposed context, the $Y$ values themselves are not exchangeable under the null hypothesis, as the predictors included in the model together with the variable under study are nuisance variables that could explain some portion of variability of $Y$. In order to address this problem, among a number of available techniques, the Freedman-Lane procedure \citep{freedman1983} has proved to provide the best control of statistical power and false positives (type 1 error) \citep{winkler2014}. This procedure proceeds as follows. If $Z$ contains all the nuisance variables previously contained in $X$, the general linear model can be rewritten as $Y = \beta X +  \gamma Z + \epsilon$. Then, instead of permuting $Y$ and extracting $\beta$, the procedure computes the residual-forming matrix $R_{Z} = 1 -  X  X^T$ and performs $N$ different permutations by computing the model $\hat{{P_N}} R_{Z}  Y = \beta X +  \gamma Z + \epsilon$ at each point, where $\hat{{P_N}}$ is the permutation operator. For a full derivation of the Freedman-Lane strategy see \emph{Winkler et al.} \citep{winkler2014}.

\subsection{False discovery rate correction for multiple comparisons}
A multiple testing problem arises by testing tens of thousands of statistical hypotheses simultaneously. Control of the family wise error rate at 5\% could be derived by extracting the maximum score from each map derived via permutation testing and by using the 95th percentile as a threshold for significance. However, in this context such a correction could be overly conservative as we are rarely interested in the exact number of vertices that reach significance. The main goal is to detect extended areas of coherent signal and therefore we can accept a maximum fixed percentage of false discoveries as provided by false discovery rate (FDR) procedures. In particular, these procedures can be applied to adjust the voxelwise p-values obtained at each vertex by computing the ratio between the number of times in which a TFCE score greater than the measured one has been obtained and the number of permutations $N$. We have found adaptive procedures such as the two-stage Benjamini-Hochberg \citep{benjamini2006} not suitable for our dataset, since it led to lower p-values and increased areas of significance, as also reported in the neuroimaging literature \citep{reiss2012}. For this reason, the original Benjamini-Hochberg (BH) \citep{benjamini1995} procedure has been employed for this work. It is important to underline that both FDR correction procedures are valid when the tested hypotheses are independent or satisfy a technical definition of dependence called positive regression dependency on a subset \citep{benjamini2001}. This condition for Gaussian data is translated into the requirement that the correlation between null voxels or between null and signal voxels is zero or positive, and for smoothed image data as those that compose a cardiac atlas, this assumption is generally considered satisfied \citep{penny2011}.

\subsection{Software}
The proposed mass univariate framework has been coded as an R package (mutools3D) which benefits from the use of vectorized operations. Matrices containing the phenotypic data and templates to visualise the 3D models are also available with the software. Linear regression assumptions must be met in order to obtain reliable inferences (for a review of them and their importance in a mass univariate setting see \emph{Appendix B}). Particularly important in this context are multicollinearity and heteroscedasticity problems which should be checked and solved for each model definition. For this latter, the R package implements mass univariate functions exploiting HC4m heteroscedascity consistent estimators \citep{CribariNeto2011}.

\section{Results}
\subsection{GWAS Replication Study}
\begin{figure}
\includegraphics[width=0.41\textwidth]{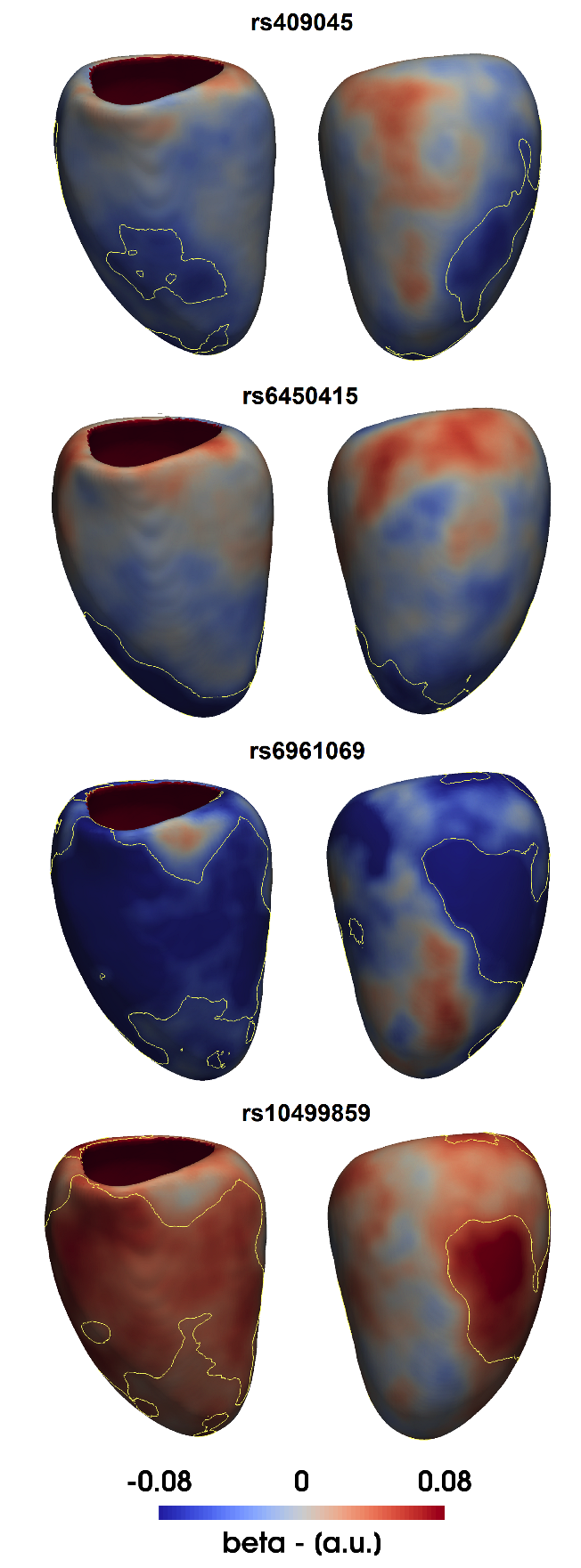}
\caption{{\bf Applying three-dimensional analysis to single nucleotide polymorphism (SNP) replication.} On the left $\beta$ coefficients are plotted on the surface of the left ventricle for the effect of rs7183401 on wall thickness (WT) adjusted for age, gender, body surface area and systolic blood pressure. Yellow contours enclose standardized regression coefficients reaching significance after multiple testing (FDR-BH) correction using the full pipeline. Standard mass univariate regression (MUR) did not provide any significant areas.  The right ventricle is shown in outline for reference. On the right, the distributions of the derived p-values adjusted for multiple testing obtained with standard MUR or the proposed pipeline are shown (the vertical line represents p-value=0.05).}
\label{fig4}
\end{figure}

As an exemplar application, 6 out of 9 exonic SNPs which have previously shown an association with LV mass in a case-control genome wide association study (GWAS), using echocardiography for phenotyping \citep{arnett2009genome}, were also identified in the UK Digital Heart Project genotypes and were assessed for replication. For each SNP, WT at each vertex in the 3D model in 1,124 healthy Caucasian subjects was tested for association with the posterior estimate of the allele frequency by a regression model adjusted for age, gender, body surface area (BSA) and systolic blood pressure (SBP). The tested SNPs are rs409045, rs6450415, rs1833534, rs6961069, rs10499859 and rs10483186 and cohort characteristics are reported in \emph{Appendix E}. Regression diagnosis through Breush-Pagan and White's test showed how the homoscedasticity assumption was violated at a large number of vertices, therefore mass univariate regression was corrected using HC4m heteroscedascity consistent estimators \citep{CribariNeto2011}. Regarding the assumption of multicollinearity, the condition number of the model matrix was 2.19 while the variance inflation factor was equal to 1.06, suggesting a very low degree of multicollinearity. All the simulations were executed on a high performance computer (Intel Xeon Quad-Core Processor (30M Cache, 2.40 GHz), 36Gb RAM), using the analysis pipeline and R package proposed in this paper (FIG. \ref{fig2}). A multiple comparisons procedure correcting for the number of vertices and the number of SNPs tested was applied by simultaneously testing in a BH FDR-controlling procedure all the TFCE-derived p-values from all the models as suggested in \citep{benjamini2005quantitative}. The number of permutations was fixed to 10,000 and simulations required less than 3 hours each. Finally, as a result of a preliminary study we conducted (full details in \emph{Appendix C}), TFCE parameters E and H were set to 0.5 and 2, as suggested in the original TFCE paper \citep{smith2009}, since this choice provides good sensitivity and specificity on a range of synthetic signals. \\
Four SNPs showed a significant association with WT as reported in FIG. \ref{fig4}. These are rs409045 (maximum regression coefficient $\bar{\beta}=-0.1$, percentage of the LV area significant $S=13\%$), rs6450415 ($\bar{\beta}=-0.11$, $S=11\%$), rs6961069 ($\bar{\beta}=-0.09$, $S=44\%$) and rs10499859 ($\bar{\beta}=0.1$, $S=41\%$). Conventional linear regression analysis using LV mass and the same model for all the SNPs did not reach significance (\emph{Appendix E}).

\begin{figure*}
\includegraphics[width=\textwidth]{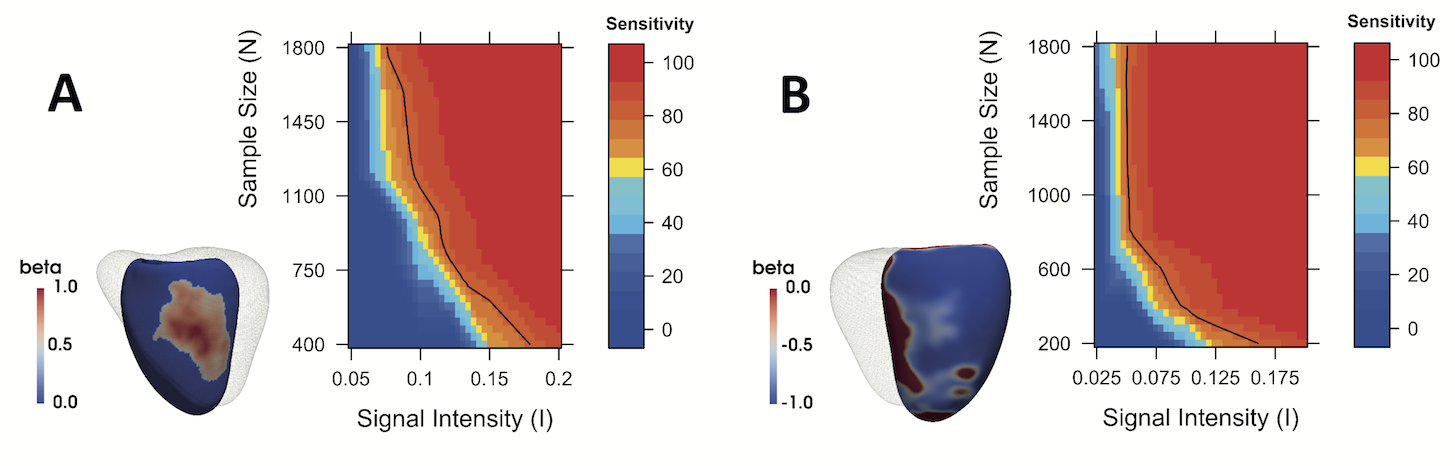}
\caption{{\bf Assessment of power using synthetic data.} Plots of our framework's sensitivity at different sample sizes $N$ and signal intensities $I$ to detect a synthetic signal on A) 10\% and B) 60\% of the LV surface. A black line on the plots indicates a threshold of 80\% sensitivity.}
\label{fig7}
\end{figure*}

\subsection{Assessment of Sensitivity, Specificity and False Discovery Rate on Synthetic Data}
Sensitivity, specificity and the rate of false discoveries of the proposed pipeline were estimated using synthetic data. A 3D model showing no correlation between WT and the posterior estimate of the allele frequency $X_{snp}$ of an non-associated SNP (rs4288653) adjusted for age, gender, BSA and SBP was used to generate background noise. A synthetic data signal was generated by summing to the WT values of each subject a term $I \; \beta \; X_{snp}$ at each vertex, where $I$ is the signal intensity and $\beta$ is a map of regression coefficients. Two contrasting $\beta$ maps (signal A and B) obtained from real clinical data were chosen and are available in \emph{Appendix }. Signal A was characterized by non-null $\beta$ coefficients covering the 10\% of total area of the LV and scaled to the (0,1] range, while signal B presented non-null regression coefficients scaled to the [-1,0) range in a more extended area covering the 60\% of the LV surface. By subsampling the number of subjects $N$ and the signal intensity $I$, different signals to be detected by the proposed standard mass univariate pipeline were obtained. The number of permutations for each simulation was fixed to 5,000 and results were linearly interpolated and plotted on the contour plots shown in FIG. \ref{fig7}. \\
Sensitivity increased at larger sample sizes $N$ and signal intensities $I$, reaching the greatest values with the most extended signal (signal B) as expected. Given the sample size of our GWAS replication study and the intensity of the associations found, these results would assign a sensitivity of 70\% for the first two discovered SNPs and more than 90\% for the other two. Moreover, the rate of false discoveries was 0 for all the results of signal A and below 5\% except for few simulations involving signal B and sample sizes greater than 1,600 (results reported in \emph{Appendix F}). This effect is due to the large synthetic signal extension, which causes TFCE to extend its support to vertices near the true signal which show the same direction of effect. This is not considered a major limitation as TFCE will not enhance clusters that originate only from noise. Finally, sensitivity, specificity and the rate of false discoveries were also computed using our pipeline without TFCE - which showed how application of TFCE provides a relevant increase of up to 50\% in sensitivity which only comes at the expense of a small decrease in specificity on large extended signals (\emph{Appendix F}). \\
Finally, we have performed a comparative study between TFCE and a standard cluster-based thresholding method as reported in the original TFCE paper on brain imaging data \citep{smith2009}, which has been implemented in the proposed R package (\emph{Appendix D}). Overall, in agreement with the neuroimaging literature, the sensitivity of the cluster-extent based thresholding method was lower than TFCE and proved to be very dependent on the cluster-forming threshold. Moreover, higher false discovery rates and lower specificities characterised cluster-extent based thresholding results in all cases when their sensitivity was comparable or greater than TFCE.

\section{Discussion}
The environmental and genetic determinants of cardiac physiology and function, especially in the earliest stages of disease, remain poorly characterized and morphological classification relies on one-dimensional metrics derived by manual image segmentation \citep{khouri20104}. In contrast, computational cardiac analysis provides precise 3D quantification of shape and motion differences between disease groups and normal subjects \citep{medrano2013atlas}. We have extended the application of these techniques by designing a general linear model framework that provides a powerful approach for modelling the relationship between phenotypic traits, genetic variation and environmental factors using high-fidelity 3D representations of the heart. By translating statistical parametric mapping techniques originally developed for brain mapping to the cardiovascular domain we exploit spatial dependencies in the data to identify coherent areas of biological effect in the myocardium. This framework also accounts for multiple testing correction at tens of thousands of vertices which is the main drawback of this class of techniques. In particular, the application of TFCE leads to a notable increase in power of the mass univariate approach at the expense of only a slight increase of the false discovery rate in  large extended signals. \\
Genetic association studies using conventional 2D imaging leave much of the moderate heritability of LV mass unexplained \citep{vasan,fox,semsarian2015}. One contribution may be the lack of phenotyping power of conventional imaging metrics, which require manual analysis and are insensitive to regional patterns of hypertrophy. Our simulations on synthetic data show that our approach has the power to detect anatomical regions associated with even small genetic effect sizes. In the reported exemplar application, we replicated the effect of four SNPs discovered in a GWAS for LV mass using a 3D WT phenotype with TFCE applied, while none of the SNPs replicated with conventional LV mass analysis. The genotype-phenotype associations that we report reflect that cardiac geometry is a complex phenotype with a highly polygenic architecture dependent on anatomical patterns of gene expression and spatially-varying adaptations to haemodynamic conditions \citep{wild2017large,srivastava2000genetic}. \\
One of the main limitations of the presented framework is that high-spatial resolution CMR is not available in all cohorts, although conventional two-dimensional images may be super-resolved to provide similar shape models \citep{oktay2017anatomically}. A second limitation is that the true association may not be linear in the model parameters and nonlinear models could better fit the data. However, the advantages favouring a general linear model are its simplicity, the ability to easily design and adjust the results for multiple factors and its wide use in biomedical statistics. A third limitation of this work is with regards to the experiments using synthetic data as we only assessed noise in our single centre population and did not generalise this to other cohorts. A general limitation of these approaches is that they do not establish causal relationships, such as the interaction between genetic variants, blood pressure and LV mass, although this may be addressed in future work by Mendelian randomisation. \\
Mass univariate approaches do not directly consider the local covariance structure of the data, however this is accounted for when Random Field Theory or permutation tests define a threshold for significant activation \citep{bronzino2014biomedical}. In the neuroimaging literature, in the context of brain-wide candidate-SNP analyses, mass univariate approaches are used more extensively than multivariate approaches as the latter are less sensitive to regional effects and they require more observations than the dimension of the response variable (i.e. number of vertices) or the use of dimensionality reduction techniques \citep{penny2011}. \\
As the methods are computationally-efficient and require no human input for phenotypic analysis, it is feasible to scale up the pipeline to larger population cohorts such as UK Biobank, which aims to investigate up to 100,000 participants using MR imaging  \citep{petersen2013}. Applying these concepts to revealing the effect of rare variants on LV geometry in participants without overt cardiomyopathy \citep{Schafer2016} and to vertex-wise genome-wide analyses also represent an interesting area of future work. In this latter context, multivariate approaches may show promise for modelling high-dimensional imaging and genetic data \citep{vounou2012sparse,liu2014review}. Finally, while we have focused on LV geometry and shape, the same approach can be applied to time-resolved vertex-wise data to create a functional phenotype for regression modelling.

\section{Conclusion}
We report a powerful and flexible framework for statistical parametric modelling of 3D cardiac atlases, encoding multiple phenotypic traits, which offers a substantial gain in power with robust inferences. We have implemented and validated the approach on both synthetic and clinical datasets, showing its suitability for detecting genotype-phenotype interactions on LV geometry. More generally, the proposed method can be applied to population-based studies to increase our understanding of the physiological, genetic and environmental effects on cardiac structure and function. 

\begin{acknowledgments}
The authors thank Dr. James Ware for advice on GWAS replication; our radiographers Ben Statton, Marina Quinlan and Alaine Berry; our research nurses Tamara Diamond and Laura Monje Garcia; and the study participants. 
\end{acknowledgments}

\section*{Funding}
The study was supported by the Medical Research Council, UK; National Institute for Health Research (NIHR) Biomedical Research Centre based at Imperial College Healthcare NHS Trust and Imperial College London, UK; British Heart Foundation grants PG/12/27/29489, NH/17/1/32725, SP/10/10/28431 and RE/13/4/30184; Academy of Medical Sciences Grant (SGL015/1006) and a Wellcome Trust-GSK Fellowship Grant (100211/Z/12/Z).

\section*{Appendices}

\appendix

\section{UK Digital Heart Project protocols}
\label{ap1}
Participants who had known cardiovascular or metabolic disease or were taking prescription medicines were excluded but simple analgesics, antihistamines, and oral contraceptives were acceptable. Female subjects were excluded if they were pregnant or breastfeeding. Standard safety contraindications to magnetic resonance imaging were applied including a weight limit of 120 kg. All subjects provided written informed consent for participation in the study, which was approved by a research ethics committee. Anthropometric measurements were collected by trained research nurses. Subjects fasted for four hours before the visit took place. Blood pressure was acquired in accordance with the guidelines of the European Society of Hypertension \citep{OBrien2002} using a calibrated oscillometric device (Omron M7, Omron Corporation, Kyoto, Japan). Five measures were taken, the first three were discarded and the last two were averaged. Subjects height (Ht) and weight (Wt) were assessed without shoes and body surface area (BSA) was calculated by the Mosteller formula: $ BSA (m^2) = \sqrt{\frac{Ht(cm) * Wt (kg)}{3600}}$.\\ 
Cardiac MR (CMR) was performed on a 1.5-T Philips Achieva system (Best, the Netherlands). To capture the whole-heart phenotype, a high-spatial resolution 3D balanced steady-state free precession cine sequence was used that assessed the left and right ventricles in their entirety in a single breath-hold (60 sections, repetition time 3.0 ms, echo time 1.5 ms, flip angle 50$^{\circ}$, field of view 320 $\times$ 320 $\times$ 112 mm, matrix 160 $\times$ 95, reconstructed voxel size 1.2 $\times$ 1.2 $\times$ 2 mm, 20 cardiac phases, temporal resolution 100 ms, typical breath-hold 20 s).  Images were stored on an open-source database (MRIdb, Imperial College London, UK) \citep{mridb}. Conventional volumetric analysis of the cine images was performed using CMRtools (Cardiovascular Imaging Solutions, London, UK) following a standard protocol \citep{schulz2013standardized}.

\section{Gauss-Markov assumptions}
\label{ap2}
In this section we review the statistical assumptions of the general linear model and their importance in a mass univariate context, so as to clarify under which conditions reliable inference can be obtained. \\
The regression coefficients $\beta$ in the general linear model under study can be obtained via ordinary least squares (OLS): $\beta= (X X^T)^{-1}  X^T y$ and  $s.e.(\beta) =  \sqrt{s^2 (X X^T)}$, where $s^2$ is the OLS estimate of the variance $\sigma^2$ of the observations. According to the Gauss-Markov theorem, the ordinary least square method will be the best linear unbiased estimator (BLUE) of the regression coefficients $\beta$ if its five assumptions are satisfied. Here we report the importance of these assumptions for the approach proposed in this work. 

\begin{enumerate}
  \item \emph{No multicollinearity in X.} Multicollinearity is present when a covariate is a linear combination (perfect) or is highly correlated (imperfect) with other covariates.  The absence of perfect collinearity is necessary to guarantee that $X$ is full rank so that $X X^T$ can be inverted when deriving the regression coefficients, assuring their uniqueness. Imperfect multicollinearity results in a reduction of the statistical efficiency of the derived regression coefficients, causing a reduction of power and ambiguous effects \citep{andrade1999}. In order to diagnose this, low pairwise correlations between predictor variables are not sufficient to exclude multicollinearity with more than two explanatory variables, but they are a necessary condition. The variance inflation factor and the condition number of the design matrix can be instead employed  to estimate the amount of variance of each regression coefficient increased because of collinearity and to detect the amount of collinearity respectively \citep{hair2009}. Taking together these three latter indices, modifications of the design matrix via omission or orthogonalization of explanatory variables can be considered to correct for multicollinearity.
   \item \emph{Random sampling of the population.} This assumption is required in order to not introduce bias in the estimates and it is guaranteed when candidates have been randomly selected to participate to the study.
  \item \emph{No endogeneity in X.} Endogeneity happens when an explanatory variable is correlated with the error term, \emph{i.e.} $E(\epsilon_i | X) \neq 0$. In general, this problem can be due to a model misspecification problem where one or more predictors have not been included into the model, to a measurement error in X that will add bias to the estimation of the regression coefficients, or to a simultaneity error when one column of X is jointly determined with Y. As endogeneity is a conceptual problem, there are no direct statistical tests to verify this assumption, hence the results should be questioned each time. In the context under study, the last source of endogeneity can be considered negligible as imaging and clinical data comes from different sources. Bias due to measurement errors should be considered and addressed in the experimental design definition. Finally, the first assumption implies that the proposed approach can only prove association and not cause-effect relationships.
  \item \emph{Linearity and additivity.} The relation between dependent and independent variables is required to be linear in the parameters $\beta$, therefore this assumption requires the model covariates to be correctly specified. More importantly it is required that the independent variables are additive, \emph{i.e.} the amount of change in Y associated with the increase of one predictor is independent of the values of the other covariates. This latter assumption guarantees the correct interpretation of the obtained regression coefficients, avoiding their overestimation. In our context, non-additivity can be addressed by defining interaction terms of the predictors. Moreover, when interpreting the derived regression coefficients, this assumption highlights again that this approach only shows the presence of associations and not cause-effect relationships.
  \item \emph{Homoscedasticity.} This assumption requires that the error variance $\sigma$ is identical across observations. It can be tested by computing a heteroscedasticity test such as the Breusch-Pagan (for linear heteroscedasticity) or the White's test (for non-linear heteroscedasticity). Heteroscedasticity causes too wide or too narrow regression coefficient standard errors, giving too much weight to certain subsets of the data when estimating the $\beta$s. Important sources of this effect rely in physiological or artifactual effects that underlie the measurements and they can be reduced by using either using weighted least squares, log transformations of the data or heteroscedasticity-consistent robust standard errors. \\
\end{enumerate}

If assumptions 1-4 are valid the estimation of the regression coefficients are unbiased and consistent, and if 5 is also valid it becomes efficient.

\begin{table*}
\centering
\begin{tabular}{lccc}
\hline
\toprule
    & \textbf{Full Cohort} ($N=1,124$)    & \textbf{Males} ($N=511$) &    \textbf{Females}($N=613$)             \\
\textbf{Age}  {[}$years${]}     & 43.4 $\pm$ 13.3 (19-77) & 43.2 $\pm$ 13.0 (19-77) & 43.5 $\pm$ 13.2 (20-77)  \\
\textbf{BSA} {[}$m^2${]}   & 1.84 $\pm$ 0.2 & 1.98 $\pm$ 0.16    & 1.72 $\pm$ 0.14         \\
\textbf{SBP} {[}$mmHg${]}   & 119.3 $\pm$ 14 & 125.0 $\pm$ 12.7    & 114.65  $\pm$ 13.2          \\
\hline
\vspace{1mm}
\end{tabular}
\caption{A summary of the 1,124 Caucasian subjects of UK Digital Heart Project at Imperial College cohort whose MRI scans has been used in this work.}
\label{tableCo}
\end{table*}

\begin{table*}
\centering
\begin{tabular}{lccc}
\hline
\toprule
 & \textbf{beta}   & \textbf{p-value} \\
\textbf{rs409045}    & 0.06 & 0.17   \\
\textbf{rs6450415}   & 0.01 & 0.75   \\
\textbf{rs1833534}   & -0.05 & 0.43  \\
\textbf{rs6961069}   & -0.01 & 0.96   \\
\textbf{rs10499859}  & 0.01 & 0.84   \\
\textbf{rs10483186}  & 0.01 & 0.74  \\
\hline
\vspace{1mm}
\end{tabular}
\caption{Regression coefficients and their related p-values of the linear association study between LVM and the posterior estimate of the allele frequency adjusted for age, gender, body surface area (BSA) and systolic blood pressure (SBP) of the presented GWAS replication study.}
\label{tableReg}
\end{table*}

\section{A study on the effect of the TFCE parameters E and H}
\begin{figure*}
\centering
 \includegraphics[scale=1]{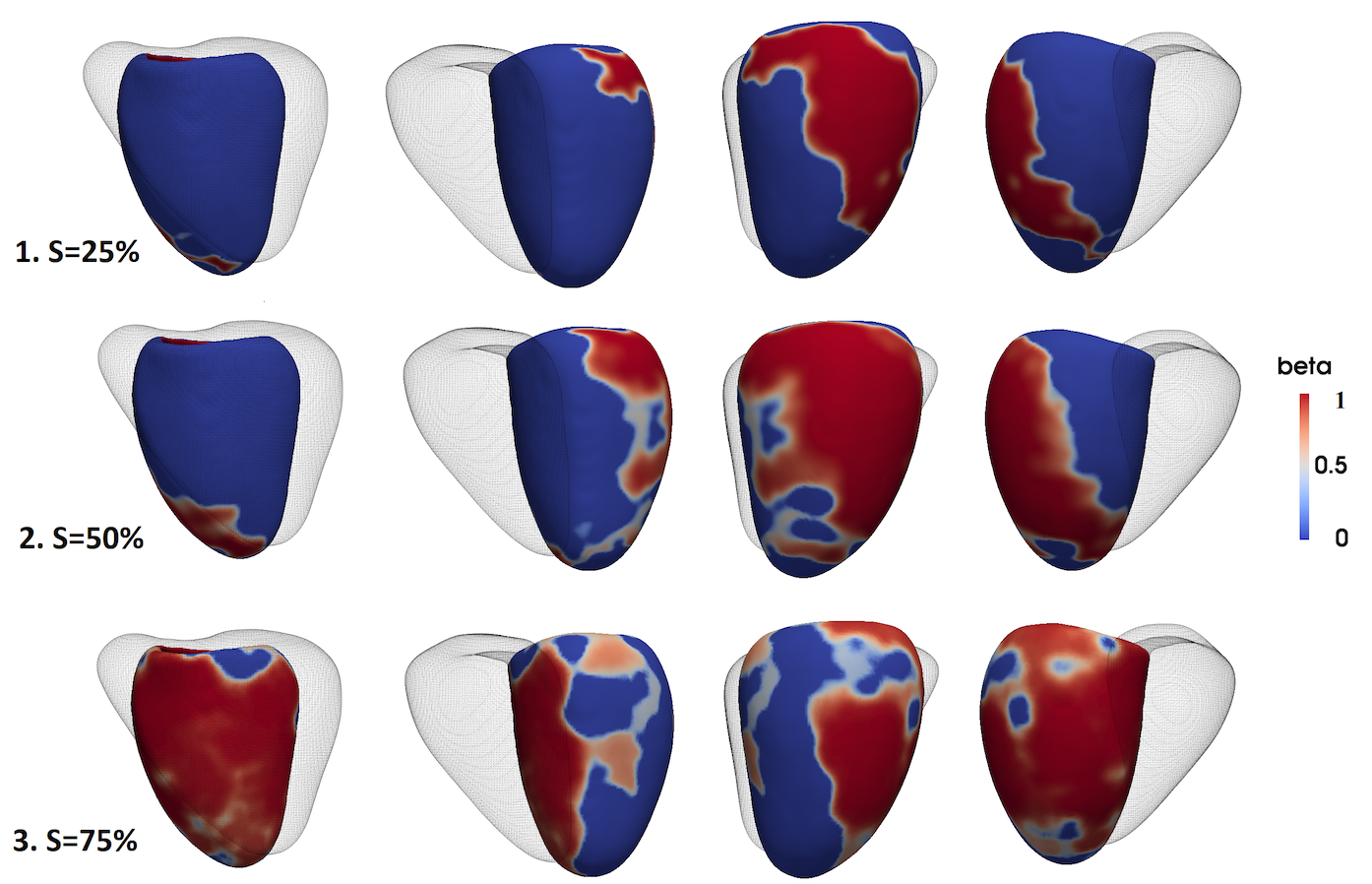}
\caption{The three regression coefficient $\beta$ maps (signal 1, 2 and 3) used to generate the synthetic signal to be detected the method proposed in this paper with different values of the TFCE parameters E and H.}
\label{signals123}
\end{figure*}

\begin{figure*}
\centering
 \includegraphics[width=0.83\textwidth]{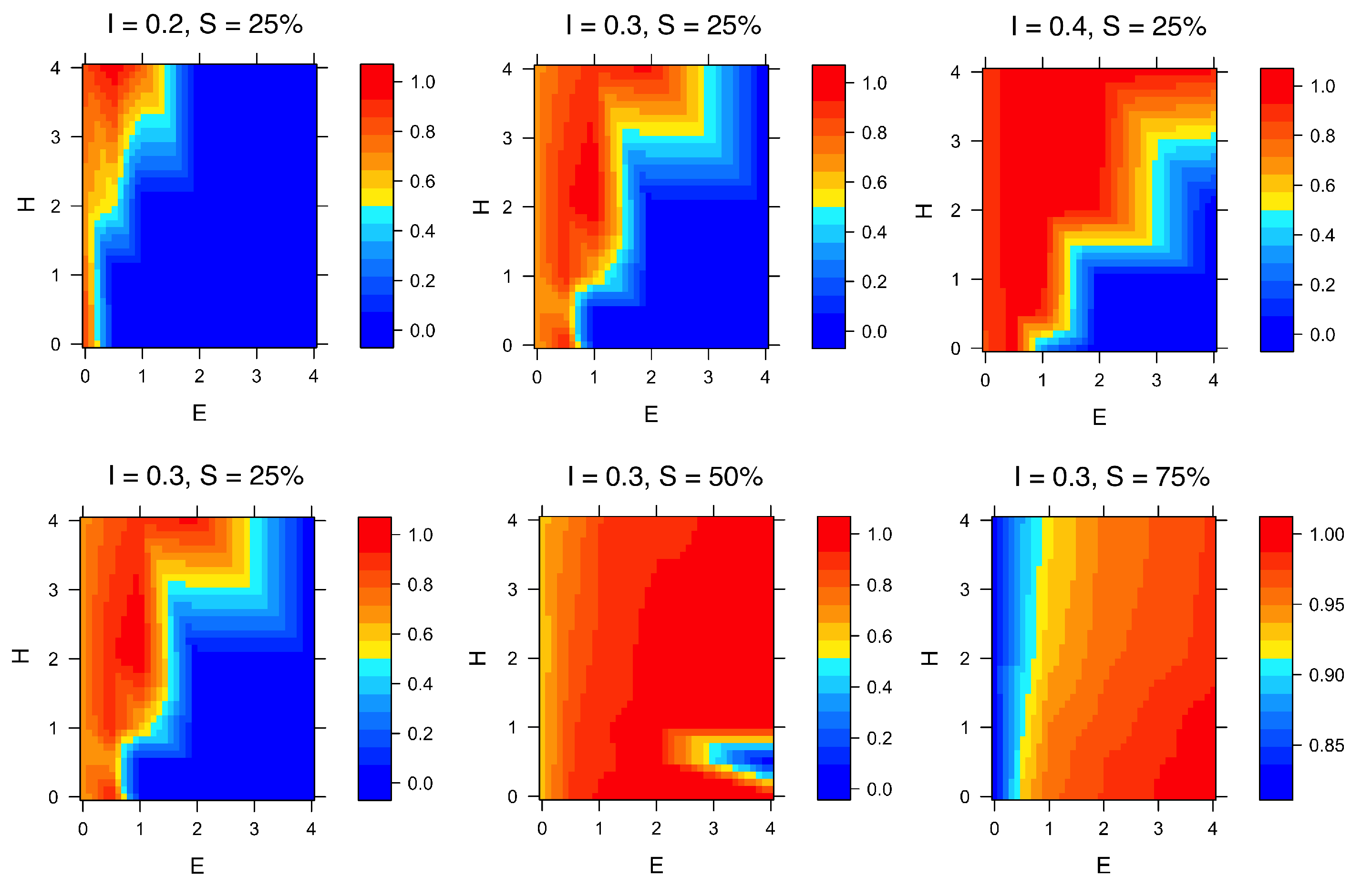}
\caption{Colour plots for the proposed framework sensitivity at different TFCE parameters E and H and for different signal intensities I and signal extension S. In each graph sensitivity values were normalized to the maximum sensitivity detected.}
\label{fig3}
\end{figure*}

The sensitivity of the proposed pipeline using different values of the TFCE parameters E and H were assessed using synthetic data. A 3D model showing no correlation between WT and the posterior estimate of the allele frequency $X_{snp}$ of a SNP (rs4288653) adjusted for age, gender, BSA and SBP was used to generate background noise. A synthetic data signal was generated by summing to the WT values of each subject a term $I \; \beta \; X_{snp}$ at each vertex, where $I$ is the signal intensity and $\beta$ is a map of regression coefficients. Three distinct $\beta$ maps (signal 1, 2 and 3) obtained from real clinical data and characterized by non-null $\beta$ coefficients scaled  to the (0,1] range were employed (FIG. \ref{signals123}). These covered the 25\%, 50\% and 75\% of total area of the LV respectively were employed together with three distinct values (0.2, 0.3, 0.4) of signal intensity I. For a given value of the signal intensity I and of the spatial extension S of the synthetic signal, five values of the parameter E and five values of the parameter H were employed by the proposed framework to detect the synthetic signal generated (a total of 25 simulations for each (I,S) couple). The number of subjects N was fixed to 80, the number of permutations for each simulation to 5,000. Sensitivity results were linearly interpolated and normalized for the maximum sensitivity obtained for each (I,S) couple and plotted on the colour plots reported in FIG. \ref{fig3}.\\
The first row of FIG. \ref{fig3} shows the sensitivity results obtained at a fixed spatial extension of the generated synthetic signal ($S=25\%$) and different signal intensities I. It can be noticed how at higher signal intensities the framework sensitivity increases and how on a small extended signal better sensitivity values are obtained when H is higher than E. The second row of FIG. \ref{fig3} shows the results obtained at a same signal intensity I when increasing the spatial extension S of the generated signal. In particular, in the bottom left figure it can be noticed how the importance of the signal intensity I is still predominant, while with the increase of the signal extension S the relative importance of the parameter E increases. In all the studied cases, the false discovery rate of the framework was always below the $5\%$ and often equal to 0$\%$, while the sensitivity was always above $99\%$.\\
Overall, in this preliminary study the values of $E=0.5$ and $H=2$ suggested in the TFCE original paper \citep{smith2009} by theoretical and empirical reasons achieved good sensitivity values. However, the performance of other combinations of E and H such as ($E=1,H=3$) show promise and will be investigated in future work.

\section{A comparison between standard cluster-extent based thresholding and TFCE}
\begin{figure*}
\centering
 \includegraphics[width=0.83\textwidth]{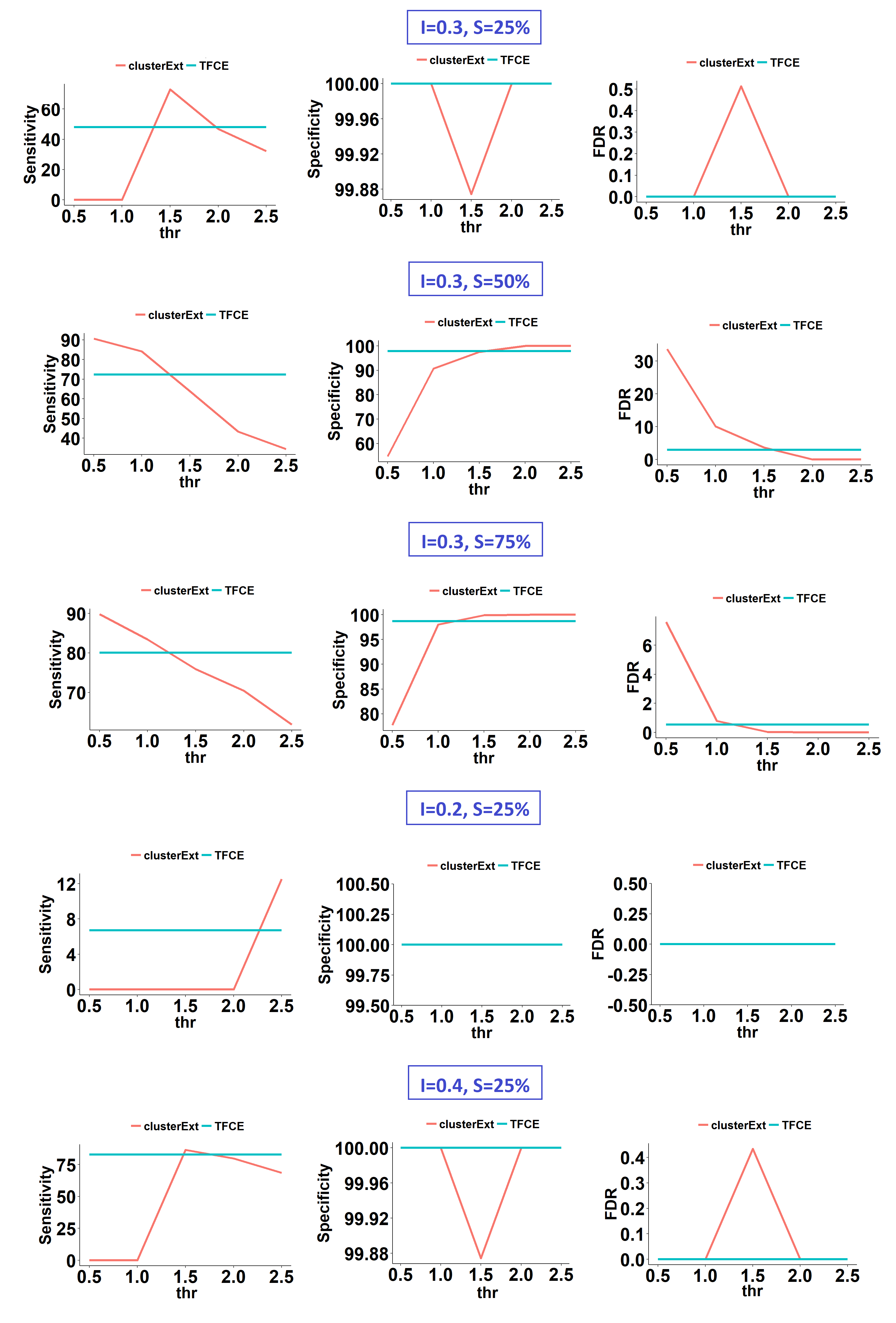}
\caption{Sensitivity, specificity and the false discovery rate (FDR) of the proposed pipeline using either cluster-extent based thresholding or TFCE at different cluster-forming thresholds (thr). At the top of each set of graphs, the intensity I and the spatial extension S of the generated synthetic signal is reported.}
\label{resClu}
\end{figure*}

A comparison between the proposed framework using TFCE and the proposed framework using a standard cluster-extend based thresholding was performed on the same synthetic data used in the previous section. The latter procedure as proposed in \citep{friston1994assessing} has been implemented in the R package developed for this work. This procedure consists of two steps. In the first one, a cluster in a statistical map obtained by mass univariate regression is defined as the group of connected vertices that have a t-statistic value greater than a user-defined threshold $h_{thr}$. Then, a second threshold $h_{\alpha}$ is computed via permutation testing as the 95th percentile of the distribution of largest cluster in each permuted map and used to to declare significant the clusters in the original statistical map that are more spatially extended than this threshold $h_{\alpha}$. Hence this method depends on the user-defined initial cluster-forming threshold $h_{thr}$. For this reason, the sensitivity, specificity and FDR of the proposed approach with TFCE parameters $E=0.5$ and $H=2$ were compared against the results obtained by the same approach using cluster-extent based thresholding with five distinct cluster-forming thresholds $h_{thr}$ ($0.5,1,1.5,2,2.5$) instead of TFCE on the five distinct signals studied in the previous section (FIG. \ref{fig3}). The number of subjects N was again fixed to 80, the number of permutations for each simulation to 5,000 and the graphs obtained are reported in FIG. \ref{resClu}. \\
The sensitivity of the cluster-extent based thresholding method proved to be very dependent on the cluster-forming threshold $h_{thr}$ and its choice had a large impact on the results. Moreover, higher FDR and lower specificity characterised cluster-extent based thresholding results when their sensitivity was comparable or greater than TFCE. These results therefore favour the use of TFCE over cluster-extent based thresholding as also proved in the brain image analysis literature \citep{smith2009}.

\section{GWAS replication supplementary data}
Table \ref{tableCo} contains a summary of the covariates employed in the regression model adopted fr GWAS replication. Table \ref{tableReg} reports the results of the linear regression models used for conventional 2D association analysis between left ventricular mass (LVM) and the posterior estimate of the allele frequency of one SNP adjusted for age, gender, body surface area (BSA) and systolic blood pressure (SBP). Even without multiple testing correction, none of the models reached significance.

\section{Sensitivity, Specificity and False Discovery Rate on Synthetic Data}
\begin{figure*}[h!]
\centering
\includegraphics[width=0.9\textwidth]{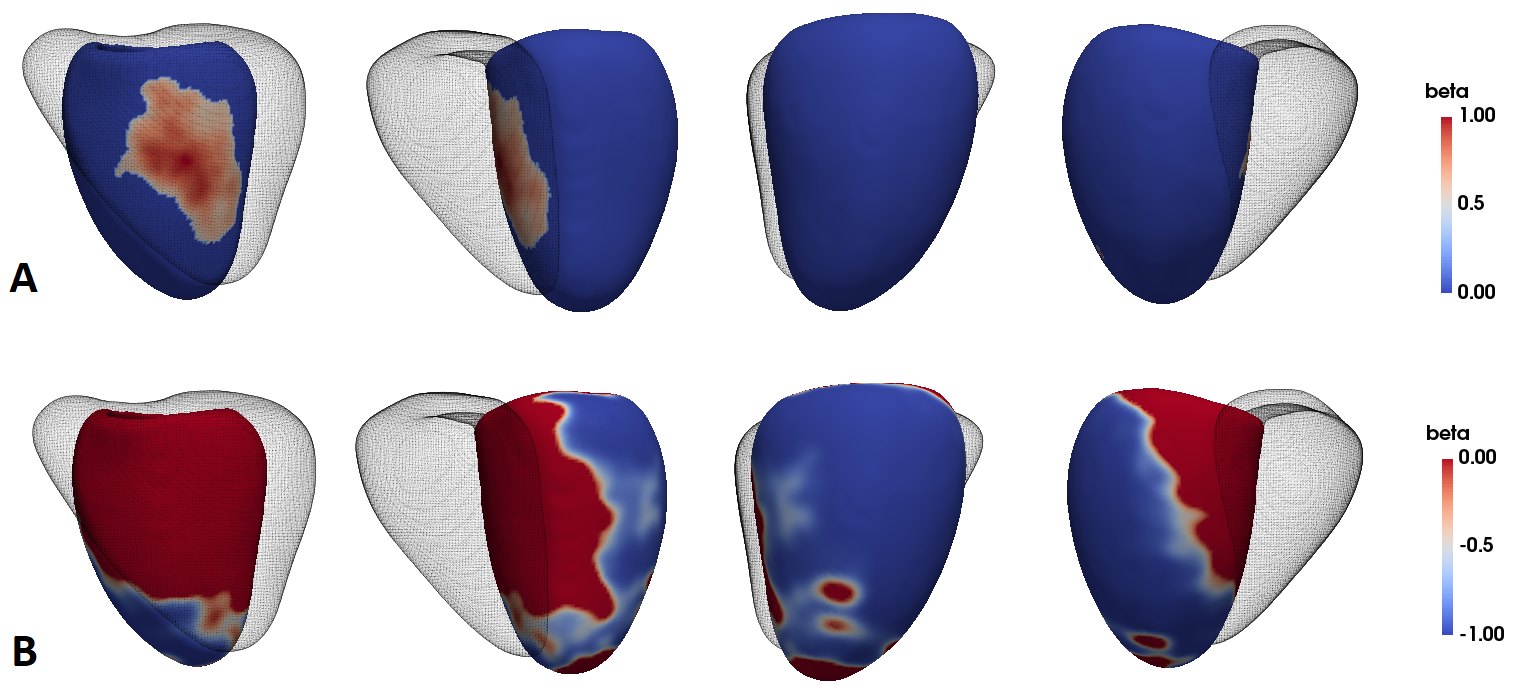}
\caption{The two $\beta$ maps used to generate the synthetic data for this experiment - signal A (first row) and B (second row).}
\label{powerGen}
\end{figure*}

\begin{figure*}[h!]
\centering
\includegraphics[width=0.9\textwidth]{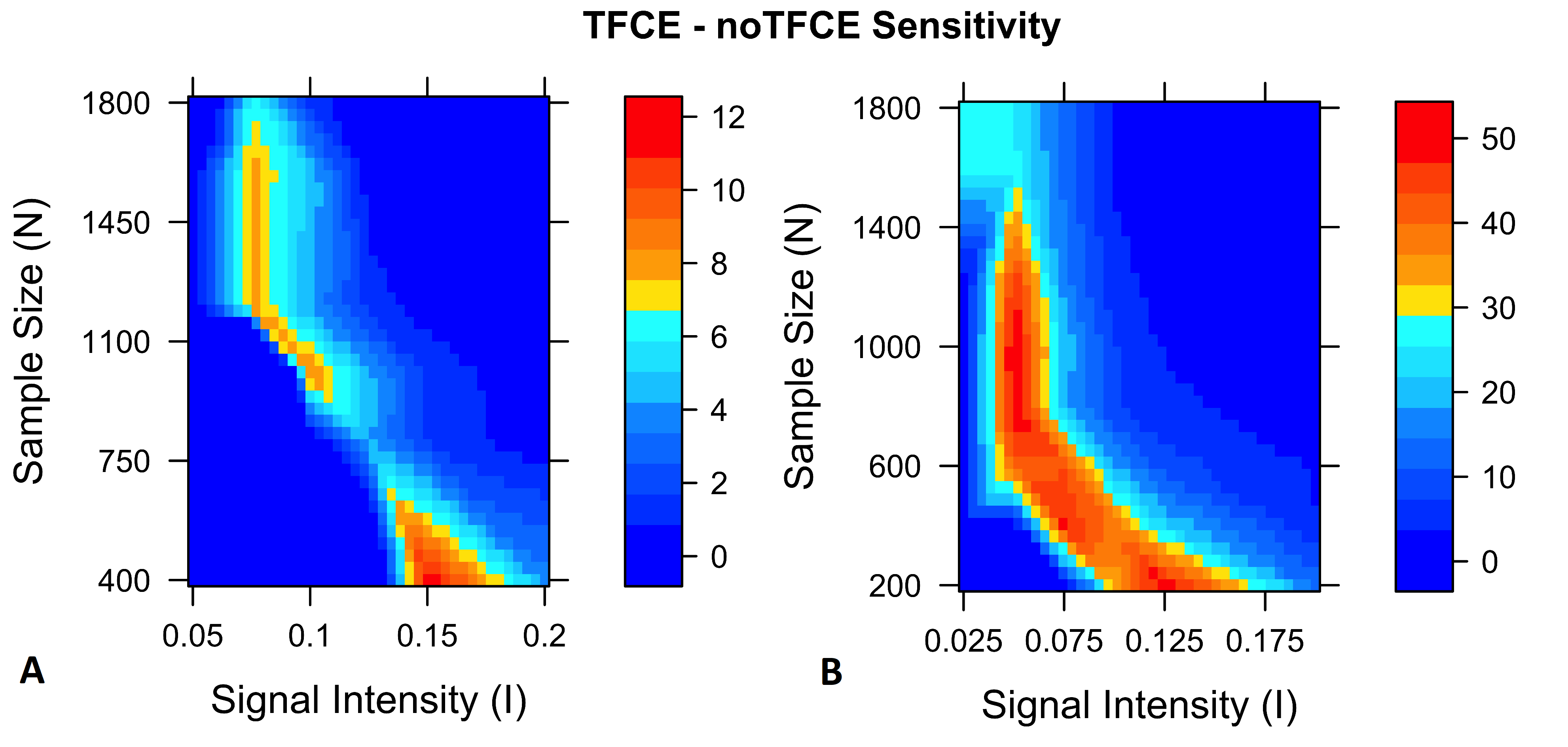}
\caption{2D maps showing the increase of sensitivity of the proposed pipeline when TFCE is applied on two different synthetic signals (A and B) at different signal intensities I and cohort dimensions (N).}
\label{diffe}
\end{figure*}

\begin{figure*}
\centering
\includegraphics[width=.7\textwidth]{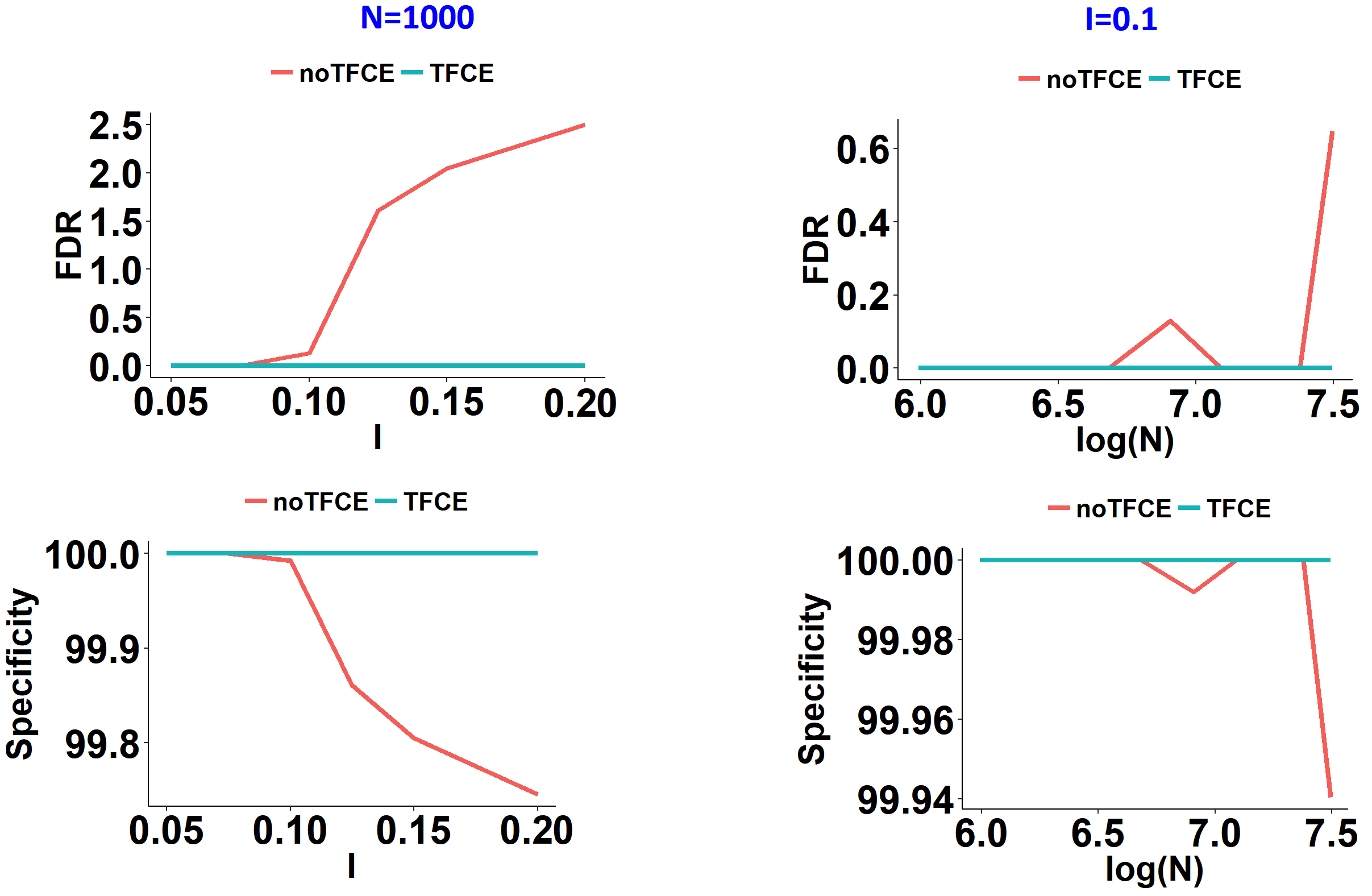}
\caption{Signal A. Rate of false discoveries and sensitivity of the proposed pipeline with or without TFCE.}
\label{specfdrA}
\end{figure*}

\begin{figure*}
\centering
\includegraphics[width=.7\textwidth]{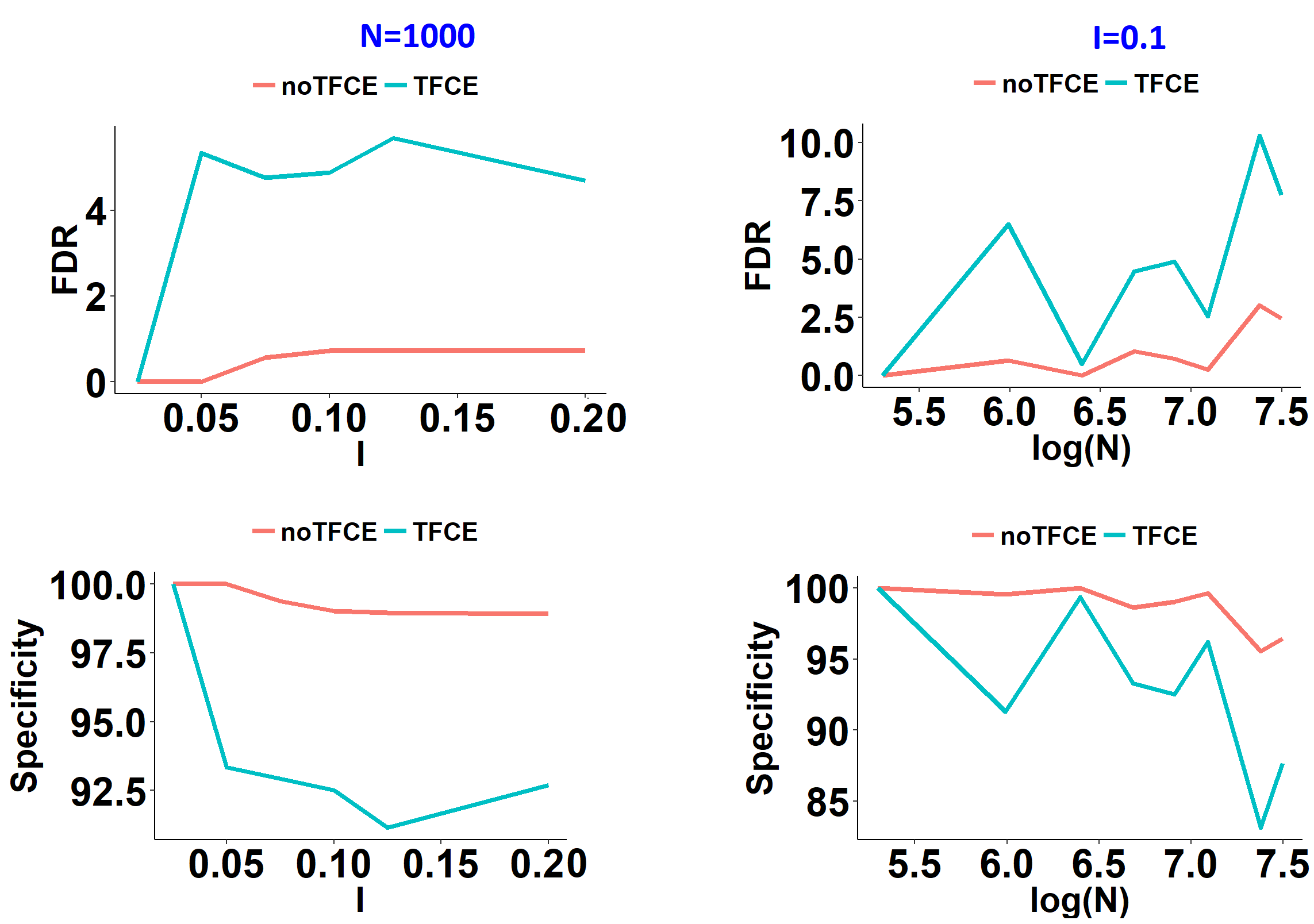}
\caption{Signal B. Rate of false discoveries and sensitivity of the proposed pipeline with or without TFCE.}
\label{specfdrB}
\end{figure*}

The two regression coefficients ($\beta$) maps used to generate the synthetic data employed in this experiment are reported in FIG. \ref{powerGen}. Signal A covers the 10\% of the left ventricle and has coefficients scaled between 0 and 1, while signal B covers the 60\% of the left ventricle and has coefficients scaled between -1 and 0. \\
FIG. \ref{diffe} reports the two maps obtained on signal A and B at different signal intensities (I) and sample sizes (N) for the difference between the sensitivity scored by the proposed pipeline using TFCE and the sensitivity scored by the same pipeline without TFCE. The increase of sensitivity provided by TFCE was higher for signal B due to its larger spatial extension as expected. The difference of sensitivities converged to zero at high I and N as also the sensitivity of the pipeline without TFCE reached 100\% sensitivity.\\
FIG. \ref{specfdrA} and FIG. \ref{specfdrB} report the detected false discovery rate (FDR) and specificity of the proposed pipeline with or without TFCE. In FIG. \ref{specfdrA} the FDR of the proposed pipeline was zero for all the simulated I and N, while it increased for the second signal (signal B - covering the 60\% of the ventricle) and exceeded 5\% only for few simulations having a sample size N greater than 1,600 (one example is reported in FIG. \ref{specfdrB}). This effect is due to the large synthetic signal extension, which causes TFCE to reward also the neighbour vertices around the true signal and it is not considered an issue since it cannot cause to declare cluster arisen from noise to be significant (as also shown in FIG. \ref{specfdrA}).\\
Overall, the application of TFCE provides a relevant increase in sensitivity which only comes at the expenses of a little decrease of specificity on largely extended signals.  

\section*{References}
\vspace{-3mm}
\bibliography{mybiblio2}

\newpage

\end{document}